\documentclass[letterpaper,11pt,aps,prd,amsmath,amssymb,floatfix,nofootinbib,reprint]{revtex4-1}
\usepackage{graphicx,color}
\usepackage[colorlinks=true,hyperfootnotes=false]{hyperref}
\usepackage{txfonts}
\usepackage{mathtools}

\newcommand{\ket}[1]{\left\lvert  #1 \right\rangle}

\makeatletter
\newsavebox\myboxA
\newsavebox\myboxB
\newlength\mylenA

\newcommand*\xoverline[2][0.75]{%
    \sbox{\myboxA}{$\m@th#2$}%
    \setbox\myboxB\null
    \ht\myboxB=\ht\myboxA%
    \dp\myboxB=\dp\myboxA%
    \wd\myboxB=#1\wd\myboxA
    \sbox\myboxB{$\m@th\overline{\copy\myboxB}$}
    \setlength\mylenA{\the\wd\myboxA}
    \addtolength\mylenA{-\the\wd\myboxB}%
    \ifdim\wd\myboxB<\wd\myboxA%
       \rlap{\hskip 0.5\mylenA\usebox\myboxB}{\usebox\myboxA}%
    \else
        \hskip -0.5\mylenA\rlap{\usebox\myboxA}{\hskip 0.5\mylenA\usebox\myboxB}%
    \fi}
\makeatother

\begin{document}

\author{M.~Albaladejo}
\affiliation{Instituto de F\'isica Corpuscular (IFIC),
             Centro Mixto CSIC-Universidad de Valencia,
             Institutos de Investigaci\'on de Paterna,
             Aptd. 22085, E-46071 Valencia, Spain}
\author{D.~Jido}
\affiliation{Department of Physics, Tokyo Metropolitan University, Hachioji 192-0397, Japan}
\author{J.~Nieves}
\affiliation{Instituto de F\'isica Corpuscular (IFIC),
             Centro Mixto CSIC-Universidad de Valencia,
             Institutos de Investigaci\'on de Paterna,
             Aptd. 22085, E-46071 Valencia, Spain}
\author{E.~Oset}
\affiliation{Instituto de F\'isica Corpuscular (IFIC),
             Centro Mixto CSIC-Universidad de Valencia,
             Institutos de Investigaci\'on de Paterna,
             Aptd. 22085, E-46071 Valencia, Spain}
             
\title{\boldmath $D^\ast_{s0}(2317)$ and $DK$ scattering in $B$ decays from BaBar and LHCb data}

\begin{abstract}
We study the experimental $DK$ invariant mass spectra of the reactions $B^+ \to \xoverline{D}^0 D^0 K^+$, $B^0 \to D^- D^0 K^+$ (measured by the BaBar Collaboration) and $B_s \to \pi^+ \bar{D}^0 K^-$ (measured by the LHCb Collaboration), where an enhancement right above the threshold is seen. We show that this enhancement is due to the presence of $D^\ast_{s0}(2317)$, which is a $DK$ bound state in the $I(J^P) = 0(0^+)$ sector. We employ a unitarized amplitude with an interaction potential fixed by heavy meson chiral perturbation theory. We obtain a mass $M_{D^\ast_{s0}} = 2315^{+12}_{-17}\ ^{+10}_{-5}\ \text{MeV}$, and we also show, by means of the Weinberg compositeness condition, that the $DK$ component in the wave function of this state is $P_{DK} = 70^{+4}_{-6}\ ^{+4}_{-8} \%$, where the first (second) error is statistical (systematic).
\end{abstract}
\maketitle

\section{Introduction}
The charmed and strange meson $D^\ast_{s0}(2317)$, with quantum numbers $I(J^P) = 0(0^+)$ was first observed in 
the isospin violating $D_s^+\pi^0$ decay channel by the BABAR Collaboration \cite{babar} and its
existence was confirmed by CLEO \cite{cleo}, BELLE \cite{belle1} and
FOCUS \cite{focus} Collaborations. Its mass, $M_{D^\ast_{s0}} \simeq 2317\ \text{MeV}$, is approximately $160\ \text{MeV}$ below the prediction of 
the successful constituent quark model for the charmed mesons of Ref.~\cite{god} (see however Refs.~\cite{Lakhina:2006fy,Segovia:2013wma,Ortega:2016mms}). 
Because of its low mass, the structure of this meson has been extensively discussed. The suggested interpretations cover a wide range: $c\bar{s}$ state \cite{dhlz,bali,ukqcd,ht,nari}, two-meson molecular 
state \cite{bcl,szc,Kolomeitsev:2003ac,Guo:2006fu,gamermann,Guo:2009ct,Cleven:2010aw,Cleven:2014oka,Faessler:2007cu,Faessler:2007gv,Flynn:2007ki}, $K-D$- mixing \cite{br},  
four-quark states \cite{ch,tera,mppr,nos} or a mixture between two-meson
and four-quark states \cite{bpp}. 

Some recent results from lattice QCD simulations \cite{sasa1,Liu:2012zya,sasa2,sasa3} have given additional support to the $DK$ molecular picture for the $D^\ast_{s0}(2317)$ state. In previous lattice studies it was studied with conventional quark-antiquark correlators, but no state with a mass below  the  $DK$ threshold was found (see {\it e.g.} \cite{Moir:2013ub}). In Refs.~\cite{sasa1,sasa2}, introducing $DK$ operators and using the 
effective range formula, a bound state (below the $DK$ threshold) with a binding energy around $40\ \text{MeV}$ was obtained. A similar result is obtained in other lattice simulations \cite{ThomasHadron}. Since the bound state appears when the $DK$ interpolators are included, a large $DK$ molecular component can be ascribed to this state, but more precise statements cannot be
done. In Ref.~\cite{Liu:2012zya} lattice QCD results for the $DK$ scattering length are obtained, and through the Weinberg compositeness condition \cite{Weinberg:1965zz,Baru:2003qq} the amount of $DK$ content in $D^\ast_{s0}(2317)$ is determined, with the result of a large fraction (around 70\%). Yet, this is done using an approximate formula for the scattering length. An improved version of this work is presented in Ref.~\cite{Yao:2015qia}, but the $DK$ probability is not mentioned there. Work along these lines is also done in Refs.~\cite{Altenbuchinger:2013vwa,Altenbuchinger:2013gaa}, using covariant chiral unitary approach. A reanalyis of the lattice spectra of Refs.~\cite{sasa1,sasa2} has been recently done in 
Ref.~\cite{sasa3}, considering the three lattice energy levels of Refs.~\cite{sasa1,sasa2} and going beyond the effective range expansion. Therefore, more quantitative analysis about the nature of the $D^\ast_{s0}(2317)$ could be performed, with the common result of a $DK$ component around $70\ \%$. 

Beyond these lattice results it is of foremost importance to have experimental data to test the internal structure of this enigmatic state. Weak decays of heavy hadrons into lighter states (that strongly interact thereafter, possibly generating resonant or bound states) offer an excellent opportunity for such a purpose \cite{Oset:2016lyh}. In the specific case of $D^\ast_{s0}(2317)$, in Ref.~\cite{Albaladejo:2015kea} it was proposed to use the $DK$ invariant mass distribution of the (so far unmeasured) decay $\bar{B}_s \to D_s^- DK$ to investigate the mass and the nature of this state.\footnote{A different decay has been also proposed in Ref.~\cite{Sun:2015uva}.} There are at least three reactions that have been actually measured that give access to the $DK$ invariant mass spectrum and which are relevant for the study of the $D^\ast_{s0}(2317)$ state. The Belle collaboration \cite{Brodzicka:2007aa} measured the decay $B^+ \to \xoverline{D}^0 D^0 K^+$, observing an enhancement right above the $DK$ threshold. The BaBar collaboration \cite{Lees:2014abp} has observed the same enhancement in the two decays $B^+ \to \bar{D}^0 D^0 K^+$ and $B^0 \to D^- D^0 K^+$. Since the reaction measured by the Belle collaboration is included in the two ones measured by the BaBar collaboration, we shall focus in this work in the latter. Finally, the LHCb collaboration \cite{Aaij:2014baa} has measured another decay, $B_s \to \pi^+ \bar{D}^0 K^-$, where an enhancement is also seen. The Belle collaboration shapes this enhancement with an {\it exponential background}, and so does the BaBar collaboration, not drawing definitive conclusions about the possible contribution of a scalar meson to this effect. On the other hand, the enhancement is partly attributed to the $D^\ast_{s0}(2317)$ state by the LHCb collaboration. In the present work, an attempt is made to explain the excess in the event distributions right above threshold as a consequence of the $D^\ast_{s0}(2317)$ state, which is associated to a bound state in the $0(0^+)$ $DK$ amplitude. We also try to quantify the $DK$ component of this state, $P_{DK}$, by means of the Weinberg compositeness condition.

On the other hand, the D0 Collaboration has recently reported on the possible existence of a new $1(0^+)$ state, $X(5568)$, in the $B_s\pi$ spectrum \cite{D0:2016mwd}. However, the LHCb Collaboration has not found any signature of this state in the same spectrum \cite{LHCb:2016ppf}. If this state actually exists, heavy quark flavor symmetry will predict a partner of it around $2.2\ \text{GeV}$ in the $D_s\pi$ channel \cite{Guo:2016nhb,Albaladejo:2016eps}, where the $D^\ast_{s0}(2317)$ has been observed. Therefore, to further constrain the analysis of the $D_s\pi$ spectrum, it is important to determine the properties of $D^\ast_{s0}(2317)$ from other sources.

The manuscript is organized as follows. After this Introduction, we set up in Sec.~\ref{sec:formalism} the formalism for the construction of the $DK$ scattering amplitude (Subsec.~\ref{subsec:amplitude}) and for the study of the aforementioned decays (Subsec.~\ref{subsec:weakdecay}). Our results are presented in Sec.~\ref{sec:results}, while conclusions are presented in Sec.~\ref{sec:conclusions}.

\section{Formalism}\label{sec:formalism}

\subsection{\boldmath $DK$ scattering amplitude}\label{subsec:amplitude}
The $D^\ast_{s0}(2317)$ is an $I(J^P)=0(0^+)$ state, and it will arise in our formalism as a $DK$ bound state. The BaBar and LHCb experiments actually measure the $D^0 K^+$ spectrum, so we need to consider the $D^0 K^+ \to D^0 K^+$ ({\it direct}, $T_d$) and $D^+ K^0 \to D^0 K^+$ ({\it crossed}, $T_c$) transition amplitudes and its relation to those with definite isospin $I$ ($I=0,1$), $T_I$. We first set our convention for isospin states,
\begin{equation}
\ket{D}_{\frac{1}{2}} = \left( \begin{array}{c} \phantom{-}\ket{D^+} \\ -\ket{D^0} \end{array} \right)~, \quad 
\ket{K}_{\frac{1}{2}} = \left( \begin{array}{c}            \ket{K^+} \\  \ket{K^0} \end{array} \right)~,
\end{equation}
which fixes the isospin eigenstates $\ket{DK}_I$,
\begin{align}
\ket{DK}_0 & = \frac{1}{\sqrt{2}} \left(  \ket{D^+ K^0} + \ket{D^0 K^+} \right)~,\\
\ket{DK}_1 & = \frac{1}{\sqrt{2}} \left(  \ket{D^+ K^0} - \ket{D^0 K^+} \right)~.
\end{align}
Assuming isospin conservation, and neglecting the $I=1$ interaction (as seen below), the amplitudes $T_{d,c}$ are given by:
\begin{align}
T_d & = \frac{T_0 + T_1}{2} \simeq \frac{T_0}{2}~,\\
T_c & = \frac{T_0 - T_1}{2} \simeq \frac{T_0}{2}~.
\end{align}

The elastic $DK$ $0(0^+)$ unitary amplitude, $T_0(s)$ (where $s$ is the center of mass energy squared), can be written as (see {\it e.g.} Refs.~\cite{Guo:2006fu,Flynn:2007ki}):
\begin{equation}
T_0(s)^{-1} = V_0(s)^{-1} - G_{DK}(s)~,
\end{equation}
where $G_{DK}(s)$ is a loop function computed from a once-subtracted dispersion relation,
\begin{align}
& 16\pi^2 G_{DK}(s) = a(\mu) +  \log\frac{m_D m_K}{\mu^2} + \frac{\Delta}{2s}\log\frac{m_D^2}{m_K^2} \nonumber  \\
& + \frac{\nu}{2s} 
\left( 
\log\frac{s-\Delta+\nu}{-s+\Delta+\nu} + 
\log\frac{s+\Delta+\nu}{-s-\Delta+\nu}
\right) \label{eq:GloopSubtracted}~, \\
& \Delta = m_D^2-m_K^2~, \quad \nu = \lambda^{1/2}(s,m_D^2,m_K^2)~.\nonumber
\end{align}
The subtraction constant $a(\mu)$ is an unknown parameter (we set the scale $\mu$ to the value $1.5\ \text{GeV}$). The $DK$ $S$-wave interaction potentials in the isospin $I$ channel, $V_I(s)$, are computed from the Heavy Meson Chiral Perturbation Theory lagrangian (\cite{Wise:1992hn,Manohar:2000dt}). Their expressions are (see {\it e.g.} Refs.~\cite{Guo:2006fu,Flynn:2007ki}):
\begin{align}
V_0(s) & = \frac{1}{4f^2} \left( -3 s + \frac{(m_D^2-m_K^2)^2}{s} + 2(m_D^2+m_K^2) \right)~, \label{eq:potential}\\
V_1(s) & = 0~.
\end{align}
It is worth noticing that the potentials are completely fixed at leading order in the combined heavy quark and chiral expansions, and hence the unitary amplitude $T_0(s)$ depends on a single parameter, $a(\mu)$.

If the amplitude has a pole at $s = M_{D^\ast_{s0}}^2$,
\begin{equation}
T_0(s) = \frac{g^2}{s-M_{D^\ast_{s0}}^2} + \cdots~,
\end{equation}
then the coupling $g^2$ can be computed as:
\begin{equation}
\frac{1}{g^2} = \frac{dT_0^{-1}(s)}{ds} = \frac{dV_0^{-1}(s)}{ds} - \frac{d G(s)}{ds}~,
\end{equation}
where the derivatives are to be evaluated at $s = M_{D^\ast_{s0}}^2$. Whence the following sum rule can be written,
\begin{equation}
1 = g^2\frac{dV^{-1}(s)}{ds} - g^2\frac{d G(s)}{ds}~.
\end{equation}
It was shown in Ref.~\cite{Gamermann:2009uq} (see also Ref.~\cite{Sekihara:2014kya} for detailed discussions), as a generalization of the Weinberg compositeness condition \cite{Weinberg:1965zz,Baru:2003qq} that the last term represents the probability $P_{DK}$ of finding the molecular $DK$ component in the $D^\ast_{s0}(2317)$ wave function,
\begin{equation}\label{eq:prob}
P_{DK}    = - g^2\frac{d G(s)}{ds}~,\quad
P_\text{other} = g^2 \frac{dV^{-1}(s)}{ds}~.
\end{equation}
Since the amplitude depends only on the parameter $a(\mu)$, then $M_{D^\ast_{s0}}$ and $P_{DK}$ are also uniquely determined by this parameter.

Finally, the $I=0$ $DK$ scattering length is defined by:
\begin{equation}\label{eq:aSLteo}
a_0 = - \frac{T_0(s_\text{th})}{8\pi \sqrt{s_\text{th}}}~,
\end{equation}
where $s_\text{th} = (m_D+m_K)^2$. In Ref.~\cite{sasa3}, this scattering length was determined, with the result:
\begin{equation}\label{eq:aSLlat}
a_0^\text{lat} = -1.5 \pm 0.5\ \text{fm}~.
\end{equation} 
This value will be used in our fits as an additional experimental input.

\subsection{\boldmath Weak decays $B \to \bar{D} D^0 K^+$ and $B_s \to \pi^+ \bar{D}^0 K^-$}\label{subsec:weakdecay}

\begin{figure}[t]
\includegraphics[height=6cm,keepaspectratio]{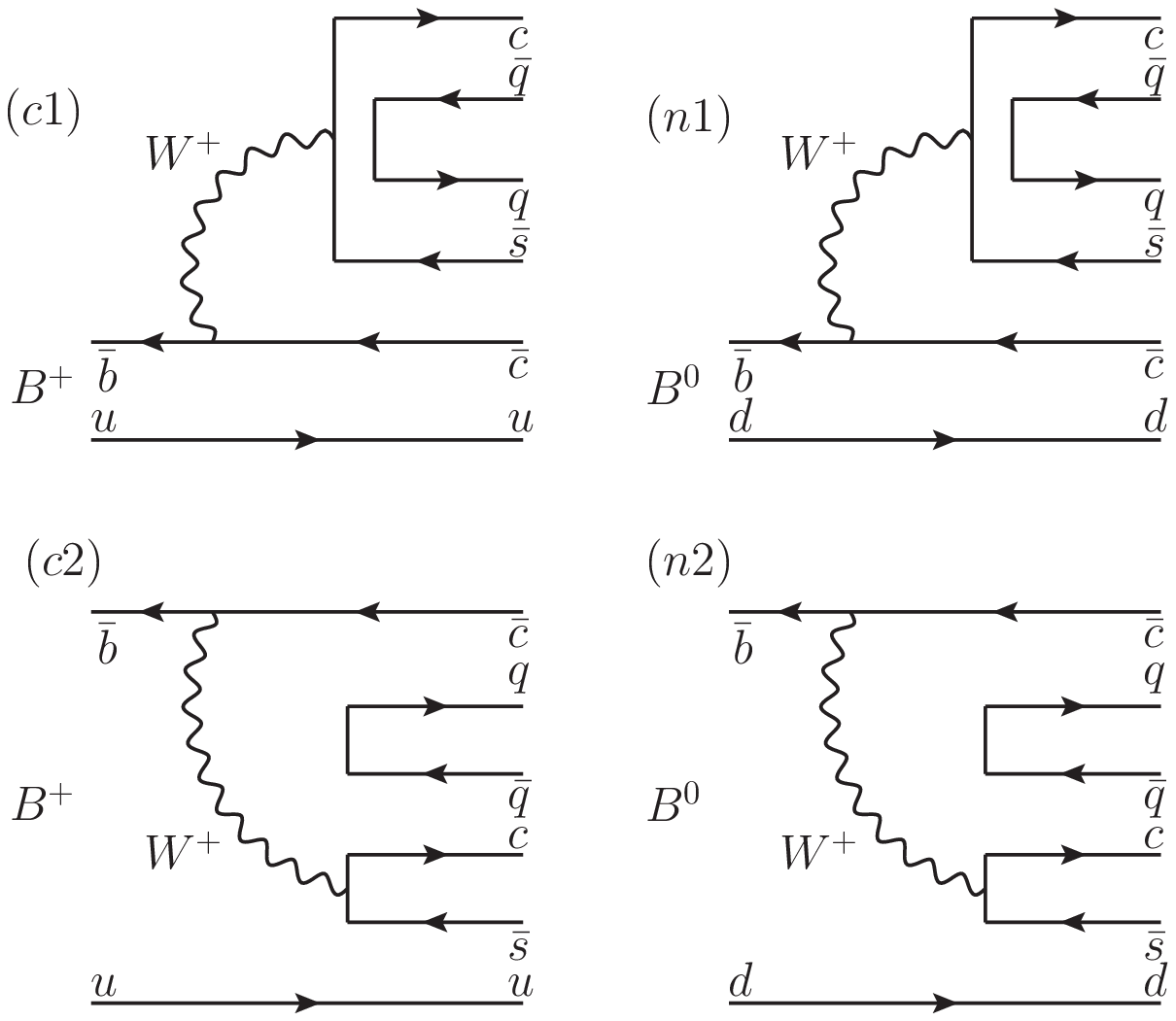}
\caption{Schematic representation at the quark level of the $B \to \xoverline{D} D^0 K^+$ decays.\label{fig:quarks}}
\end{figure}

\begin{figure}[t]
\includegraphics[height=4.5cm,keepaspectratio]{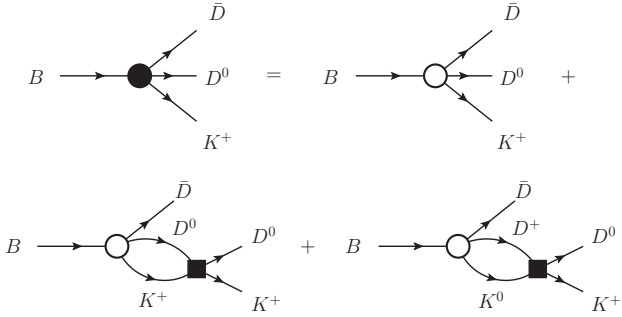}
\caption{Resummation of diagrams for $B \to \xoverline{D}D^0 K^+$ decays taking into account the $I=0$ $S$-wave $DK$ rescattering effects.\label{fig:bubbles}}
\end{figure}

We want to study the two processes $B \to \xoverline{D} D^0 K^+$, where $B$ ($\xoverline{D}$) can refer to $B^+$ ($\xoverline{D}^0$) or $B^0$ ($D^-$). The process is mediated by the weak decays $\bar{b} \to W^+ \bar{c} \to (c\bar{s})\bar{c}$. In order to have a three-meson final state, an extra $q\bar{q}$ pair must be created {\it ex vacuo}. Since a $D^0 K^+$ state must be present in the final state, it can be produced either directly or through the transition $D^+ K^0 \to D^0 K^+$ after a $D^+ K^0$ pair appears in the hadronization. There are two diagrams that contribute to each of the decays ($B^+$ or $B^0$), as depicted in Fig.~\ref{fig:quarks}. In diagram $(c1)$ the $\bar{s}c$ pair produced by the $W$ decay hadronizes together with a $\bar{q}q$ pair into a two-meson final state, and the remaining $\bar{c}u$ produces a $\xoverline{D}^0$. In diagram $(c2)$, with the topology of internal emission \cite{Chau:1982da}, the $\bar{q}q$ is {\it inserted} between the $\bar{c}c$ pair, and the $\bar{s}u$ one gives rise to a $K^+$. An analogous discussion can be applied to diagrams $(n1)$ and $(n2)$ for the case of $B^0$ decay. The $\bar{c}d$ pair in $(n1)$ gives now a $D^-$ and the $\bar{s}d$ pair in $(n2)$ gives a $K^0$ . To see the specific two-meson states that arise in the hadronization of a given quark-antiquark pair plus an extra $\bar{q}q$ pair, we introduce the following quark--anti-quark matrix $M$,
\begin{equation}
M = v \bar{v} = \left( \begin{array}{c} u \\ d \\ s \\ c \end{array} \right) \left( \begin{array}{cccc} \bar{u} & \bar{d} & \bar{s} & \bar{c} \end{array} \right)
= \left( \begin{array}{cccc} 
u\bar{u} & u\bar{d} & u\bar{s} & u\bar{c} \\
d\bar{u} & d\bar{d} & d\bar{s} & d\bar{c} \\
s\bar{u} & s\bar{d} & s\bar{s} & s\bar{c} \\
c\bar{u} & c\bar{d} & c\bar{s} & c\bar{c} \\
\end{array}\right)~,
\end{equation}
which fulfils:
\begin{equation}
M^2 = (v \bar{v})(v \bar{v}) = v (\bar{v} v) \bar{v} = \left( \bar{u}u + \bar{d}d + \bar{s}s + \bar{c}c \right) M~.
\end{equation}
The first factor in the last equality represents the $\bar{q}q$ creation. This matrix $M$ is in 
correspondence with the meson matrix $\phi$:
\begin{equation}
\phi = \left( \begin{array}{cccc} 
\frac{\eta}{\sqrt{3}} + \frac{\pi^0}{\sqrt{2}} + \frac{\eta'}{\sqrt{6}} & \pi^+ & K^+ & \bar{D}^0 \\
\pi^- & \frac{\eta}{\sqrt{3}} - \frac{\pi^0}{\sqrt{2}} + \frac{\eta'}{\sqrt{6}} & K^0 & D^- \\
K^- & \bar{K}^0 & \frac{\sqrt{2}\eta'}{\sqrt{3}}-\frac{\eta}{\sqrt{3}} & D_s^- \\
D^0 & D^+ & D_s^+ & \eta_c
\end{array}\right)~.
\end{equation}
The hadronization of the $\bar{s}c$ and the $\bar{c}c$ proceed through the matrix elements $(M^2)_{43}$ and $(M^2)_{44}$, respectively, of the $M^2$ matrix. The resulting two-meson states are then given by the same matrix elements of the $\phi^2$ matrix, namely: 
\begin{align}
(\phi^2)_{43} & = K^+ D^0 + K^0 D^+ + \cdots~,\\
(\phi^2)_{44} & = D^0 \xoverline{D}^0 + D^+ D^- + \cdots~.
\end{align}
We have retained only the terms that are relevant for the processes under consideration. Thus, in a primary step, we have in $(c1)$ $\bar{D}^0 (D^0 K^+ + D^+ K^0)$, in $(c2)$ $K^+ (D^0 \bar{D}^0 + D^+ D^-)$, in $(n1)$ $D^- (D^0 K^+ + D^+ K^0)$, and in $(n2)$ $K^0 (D^0 \bar{D}^0 + D^+ D^-)$. These configurations can be also obtained by regarding $\bar{q}q$ in Fig.~\ref{fig:quarks} as $\bar{u}u$ and $\bar{d}d$ with the same weight.

The mechanisms in Fig.~\ref{fig:quarks} give the {\it bare} vertices for the weak decays $B \to \bar{D} D^0 K^+$ and $B \to \bar{D} D^+ K^0$, and these bare vertices should be renormalized by the strong interactions among quarks. However, neither the weight of diagram $(c1)$, equal to $(n1)$, nor that of the diagram $(c2)$, equal to $(n2)$, are known, and hence we assign them a (constant) value $\gamma_1$ and $\gamma_2$, respectively. After this bare interaction takes place, the $DK$ pairs are allowed to interact, as shown in Fig.~\ref{fig:bubbles}. Let us denote by $\Gamma_{B \to \bar{D} D^0 K^+}$ the {\it full} amplitudes. Performing the summation shown in Fig.~\ref{fig:bubbles}, these are expressed as:
\begin{align}
\left( \begin{array}{c} \Gamma_{B^+ \to \bar{D}^0 D^0 K^+} \\ \Gamma_{B^0 \to      D^-  D^0 K^+} \end{array} \right) & = 
\left( \begin{array}{c} \gamma_1 + \gamma_2 \\ \gamma_1 \end{array} \right) + \nonumber \\
& \left( \begin{array}{cc} \gamma_1 + \gamma_2 & \gamma_1 \\
                         \gamma_1  & \gamma_1 + \gamma_2 \end{array} \right) G_{DK}
\left( \begin{array}{c} T_d \\ T_c \end{array} \right)
\end{align}
Now, taking into account that $T_d = T_c = T_0/2$ and the relation between $T_0$, $V_0$ and $G_{DK}$, these relations are written simply as:
\begin{align}
\Gamma_{B^+ \to \bar{D}^0 D^0 K^+} & = +\frac{\gamma_2}{2} + \left( \gamma_1 + \frac{\gamma_2}{2} \right) \frac{T_0}{V_0} = +K \left( 1 + \beta \frac{T_0}{V_0} \right)~,\label{eq:Amplis1}\\
\Gamma_{B^0 \to      D^-  D^0 K^+} & = -\frac{\gamma_2}{2} + \left( \gamma_1 + \frac{\gamma_2}{2} \right) \frac{T_0}{V_0} = -K \left( 1 - \beta \frac{T_0}{V_0} \right)~,\label{eq:Amplis2}
\end{align}
with $K = \gamma_2 / 2$ and $\beta = 1 + 2\gamma_1/\gamma_2$. The parameter $K$ is irrelevant, since it will be absorbed in a global normalization constant, and we are thus left with a single relevant parameter, $\beta$. Furthermore, our fits to the experimental data, to be discussed in more detail below, will prefer solutions with $\beta \gg 1$, which, in turn, makes the parameter $\beta$ also irrelevant, since it is again absorbed in a global normalization constant.\footnote{Since the amplitude is of the form $1 + \beta T_0/V_0$, the last term dominates for $\beta \gg 1$ unless $T_0/V_0$ has a zero, which is not the case here.} This means that the diagrams $(c1)$ and $(n1)$ in Fig.~\ref{fig:quarks} are dominant. This is an interesting empirical support for the general rule that the diagrams of external emission are color favoured and dominate the processes \cite{Chau:1982da}.

\begin{figure*}
\includegraphics[height=3.cm,keepaspectratio]{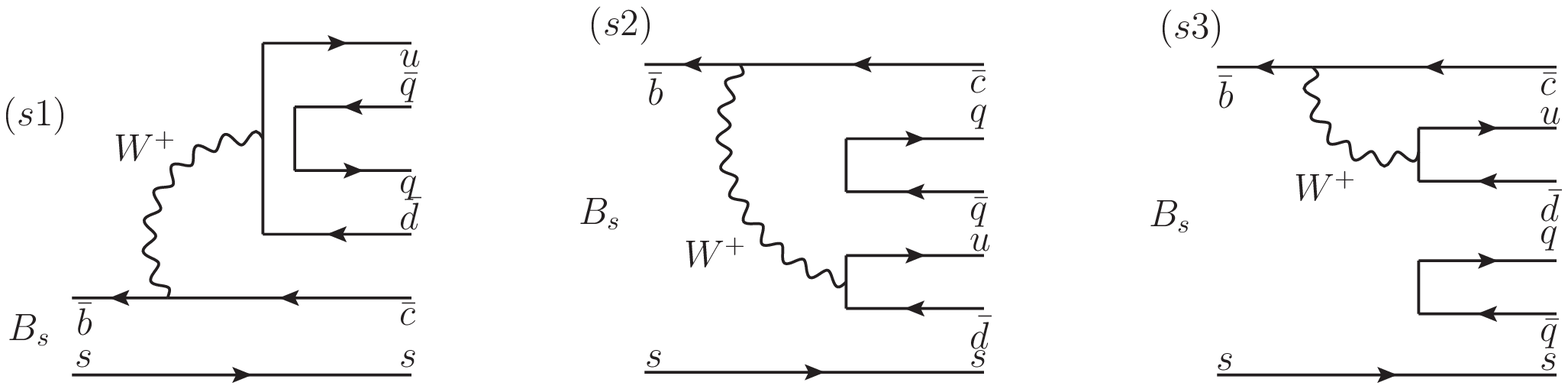}
\caption{Schematic representation at the quark level of the $B_s \to \pi^+ \xoverline{D}^0 K^-$ decays.\label{fig:Bs_decays}}
\end{figure*}
A completely analogous procedure can be taken over the reaction $B_s \to \pi^+ \xoverline{D}^0 K^-$ (with the obvious replacements), for which the relevant diagrams are depicted in Fig.~\ref{fig:Bs_decays}. The amplitude is written also as Eqs.~\eqref{eq:Amplis1} and \eqref{eq:Amplis2},
\begin{equation}\label{eq:Amplis3}
\Gamma_{B_s \to \pi^+ \bar{D}^0 K^-} = K' \left( 1 + \beta' \frac{T_0}{V_0} \right)~,
\end{equation}
and, also here, the parameters $K'$ and $\beta'$ will turn out to be irrelevant.

The experimental $DK$ invariant mass spectra in the reactions under study certainly contain contributions other than the one stemming from $DK$ with $0(0^+)$ quantum numbers, such as non-resonant background and other resonances. The full spectra, denoted here with $N_{B^+}$, $N_{B^0}$ and $N_{B_s}$ for the $B^+ \to D^- D^0 K^+$, $B^0 \to \xoverline{D}^0 D^0 K^+$, $B_s \to \pi^+ \xoverline{D}^0 K^-$ decays, respectively, are thus parameterized as follows:
\begin{widetext}
\begin{align}
N_{B^+}(E) & = p_{\bar{D}^0} p_K \left( 
\mathcal{N}_{A}^{(+)}  \left\lvert \Gamma_{B^+ \to \bar{D}^0 D^0 K^+} \right\rvert^2 + 
\mathcal{N}_{D^\ast_{s1}}^{(+)} p_K^2 \left\lvert \Gamma_{D^\ast_{s1}} \right\rvert^2 + 
\mathcal{N}_{D^\ast_{s2}}^{(+)} p_K^4 \left\lvert \Gamma_{D^\ast_{s2}} \right\rvert^2 + 
\mathcal{N}_{B}^{(+)} \left\lvert \Gamma_{B^+} \right\rvert^2 \right)~,\\
N_{B^0}(E) & = p_{D^-} p_K \left( 
\mathcal{N}_{A}^{(0)} \left\lvert \Gamma_{B^0 \to D^- D^0 K^+} \right\rvert^2 + 
\mathcal{N}_{D^\ast_{s1}}^{(0)} p_K^2 \left\lvert \Gamma_{D^\ast_{s1}} \right\rvert^2 + 
\mathcal{N}_{D^\ast_{s2}}^{(0)} p_K^4 \left\lvert \Gamma_{D^\ast_{s2}} \right\rvert^2 + 
\mathcal{N}_{B}^{(0)} \left\lvert \Gamma_{B^0} \right\rvert^2 \right)~,\\
N_{B_s}(E) & = p_{\pi^+} p_K \left( 
\mathcal{N}_{A}^{(s)} \left\lvert \Gamma_{B_s \to \pi^+ \bar{D}^0 K^-} \right\rvert^2 + 
\mathcal{N}_{D^\ast_{s2}}^{(s)} p_K^4 \left\lvert \Gamma_{D^\ast_{s2}} \right\rvert^2 + 
\mathcal{N}_{B}^{(s)} \left\lvert \Gamma_{B_s} \right\rvert^2 \right)~,
\end{align}\end{widetext}
where the different momenta involved are defined as:
\begin{align}
p_{\pi^+}(s)    & = \frac{\lambda^{1/2}(M_{B_s}^2,m^2_{\pi^+},s)}{2M_{B_s}}~,\\
p_{\bar{D}}(s)  & = \frac{\lambda^{1/2}(M_{B}^2,m^2_{\bar{D}},s)}{2M_{B}}~,\\
p_K(s)          & = \frac{\lambda^{1/2}(s,m^2_{D^0},m^2_{K^+})}{2\sqrt{s}}~.
\end{align}
The background contributions are parameterized by means of smooth energy functions,
\begin{align}
\left\lvert \Gamma_{B^+} \right\rvert^2  = \left\lvert \Gamma_{B^0} \right\rvert^2 & = p_{\bar{D}}^a\ p_K^b~,\\
\left\lvert \Gamma_{B_s} \right\rvert^2 & = p_{\pi^+}^{a'}\ p_K^{b'}~,
\end{align}
where the parameters $a,b,a'$ and $b'$ are free. The contributions from resonances other than the $D^\ast_{s0}(2317)$ are included in $\Gamma_{D^\ast_{sJ}}$. These functions are parameterized with energy dependent width Breit-Wigner functions, as done in the experimental analyses \cite{Lees:2014abp,Aaij:2014baa}. To avoid the proliferation of free parameters, the masses of the resonances included in our analysis are fixed to those given by the experimental collaborations, namely (all values in MeV) $M_{D^\ast_{s1}(2700)} = 2699 \pm 10$, $\Gamma_{D^\ast_{s1}(2700)} = 127 \pm 22$, $M_{D^\ast_{s2}(2573)} = 2568.39 \pm 0.39$ and $\Gamma_{D^\ast_{s2}(2573)} = 16.9 \pm 0.75$, where we have added in quadratures the statistical and systematic errors given by the collaborations. The normalization constants $\mathcal{N}_X^{(i)}$ are in principle free parameters.

\section{Results}\label{sec:results}

The experimental information at our disposal comprises the three event distributions for the decays $B^+ \to D^- D^0 K^+$ and $B^0 \to \xoverline{D}^0 D^0 K^+$, from the BaBar collaboration \cite{Lees:2014abp}, and $B_s \to \pi^+ \xoverline{D}^0 K^+$, from the LHCb collaboration \cite{Aaij:2014baa}, together with the result for the $0(0^+)$ $DK$ scattering length calculated in lattice simulations \cite{sasa3}, shown in Eq.~\eqref{eq:aSLlat}.\footnote{The scattering length is computed by means of Eq.~\eqref{eq:aSLteo}, and the value in Eq.~\eqref{eq:aSLlat} is included as an extra experimental point in our $\chi^2$ function. However, no significant differences are found if the scattering length is not fitted, although its inclusion in the fits improves the error estimation of the parameters and derived quantities.} For the $B_s \to \pi^+ \bar{D}^0 K^-$ spectrum, we fit the data up to $\sqrt{s} = 2.8\ \text{GeV}$, since at that energy starts the contribution from another resonance, $D^\ast_{sJ}(2860)$. In the $B \to \bar{D} D^0 K^+$ spectra, to have a more constrained fit, our background contributions are fitted to the background given in the experimental analysis of the BaBar collaboration \cite{Lees:2014abp}, by including an additional appropriate piece in the $\chi^2$ function to be minimized. The background is fitted in the whole range available for $\sqrt{s}$, while the signal data are fitted only up to $\sqrt{s} = 3\ \text{GeV}$, where the contribution of the $D^\ast_{s1}(2700)$ resonance is already small. Furthermore, in these two decays the contribution from the $D^\ast_{s2}$ is quite small, so we fix the value of the normalization constants $\mathcal{N}^{(0)}_{D^\ast_{s2}}$ and $\mathcal{N}^{(+)}_{D^\ast_{s2}}$ so as to reproduce the result given by the BaBar collaboration.

Before presenting our results, we first discuss the error estimation performed in this work. For each quantity displayed in this manuscript, the first (second) error shown is statistical (systematic). Statistical errors represent $1\sigma$ confidence intervals, and are estimated by Monte Carlo resampling of the experimental data \cite{Press:1992zz}. They also take into account the uncertainties in the masses of the $D^\ast_{sJ}$ resonances included in the spectra. The systematic errors are estimated by performing two variations in our theoretical approach. First, we consider the influence of higher orders in the potential [Eq.~\eqref{eq:potential}],
\begin{equation}\label{eq:mod1}
V_0(s) \longrightarrow V_0(s) + h (s-s_0)/s_0~,
\end{equation}
where the parameter $h$ is free. In principle, there could be also an additional term independent of $s$, but it can be absorbed, as we have checked, by a renormalization of the subtraction constant, $a(\mu)$, in the loop function [Eq.~\eqref{eq:GloopSubtracted}]. For the same token, we can take, for convenience, $s_0 = M_{D^\ast_{s0}}^2$. A potential of the type of Eq.~\eqref{eq:mod1} can account for some missing channels, which demand an energy dependent effective potential \cite{Aceti:2014ala}. Actually, in general, and as we shall see in our results, $P_{DK} \leqslant 1$, indicating missing channels. This also means that the $D^\ast_{s0}(2317)$ state can be formed directy (and not through the $DK$ rescattering) in the $B$ decay reactions. For this reason, as done in Ref.~\cite{Sun:2015uva} we consider in Eqs.~\eqref{eq:Amplis1},~\eqref{eq:Amplis2} and \eqref{eq:Amplis3} the modification:
\begin{equation}\label{eq:mod2}
\Gamma_{B^+ \to \bar{D}^0 D^0 K^+} \to \Gamma_{B^+ \to \bar{D}^0 D^0 K^+} + \frac{C_{B^+}}{s-M^2_{D^\ast_{s0}}}~,
\end{equation}
and analogously for the other two amplitudes. Both modifications [Eqs.~\eqref{eq:mod1} and \eqref{eq:mod2}] are considered separately, and the new parameters ($h$, $C_{B^+}$, \ldots) are allowed to vary together with the original ones. The contributions stemming from these new parameters is relatively small. For each quantity quoted in this work, the difference between the value obtained with the new fit and the central fit gives the systematic error.

\begin{figure*}[t]
\begin{tabular}{cc}
\includegraphics[height=5.0cm,keepaspectratio]{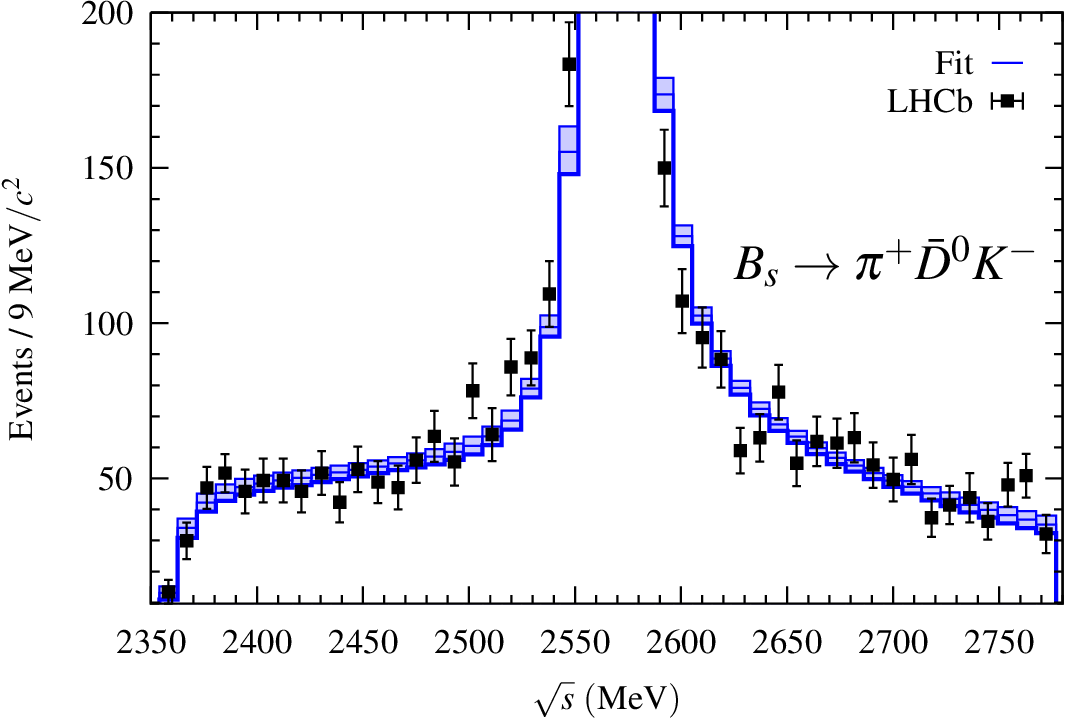} &
\includegraphics[height=5.0cm,keepaspectratio]{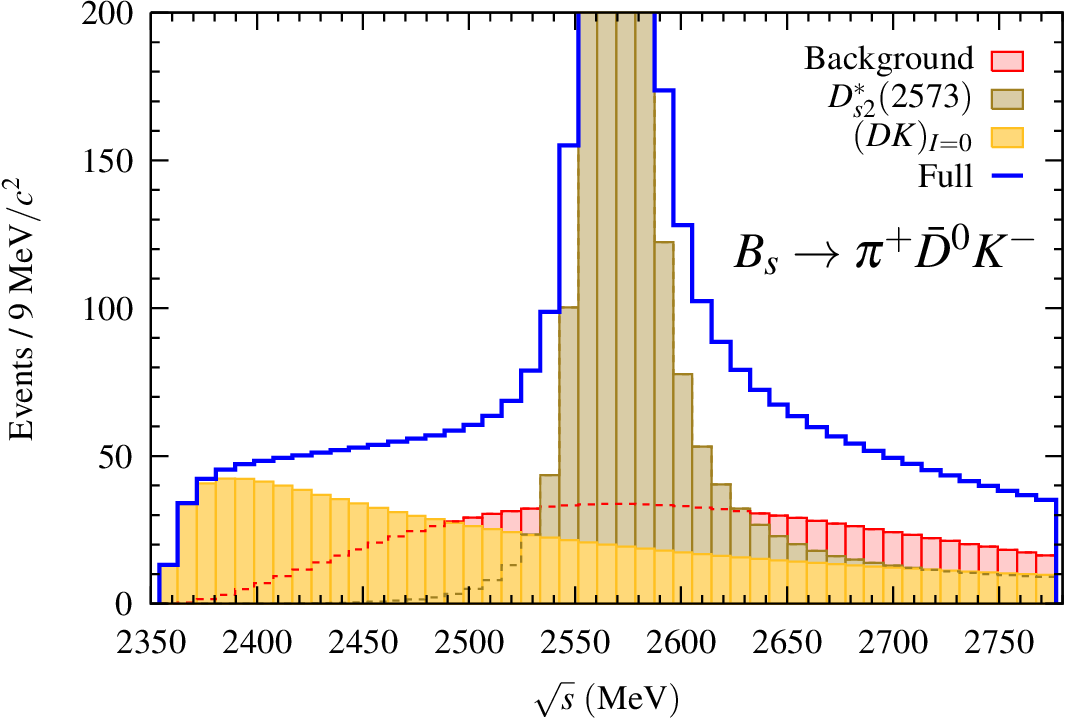} \\
\includegraphics[height=5.0cm,keepaspectratio]{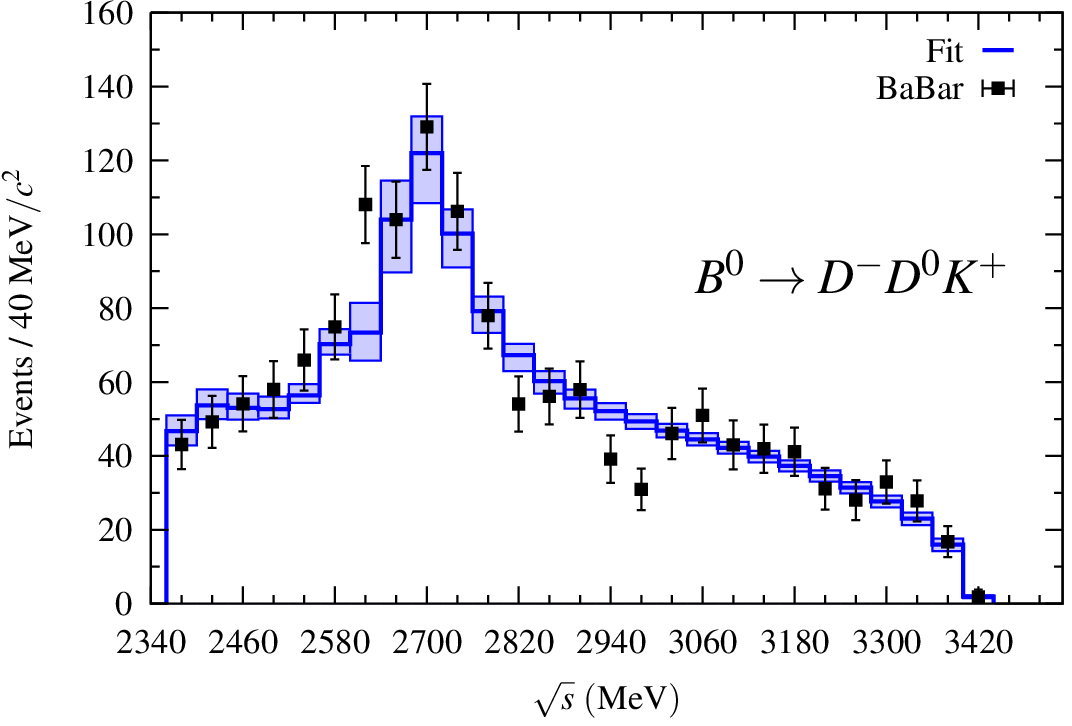} &
\includegraphics[height=5.0cm,keepaspectratio]{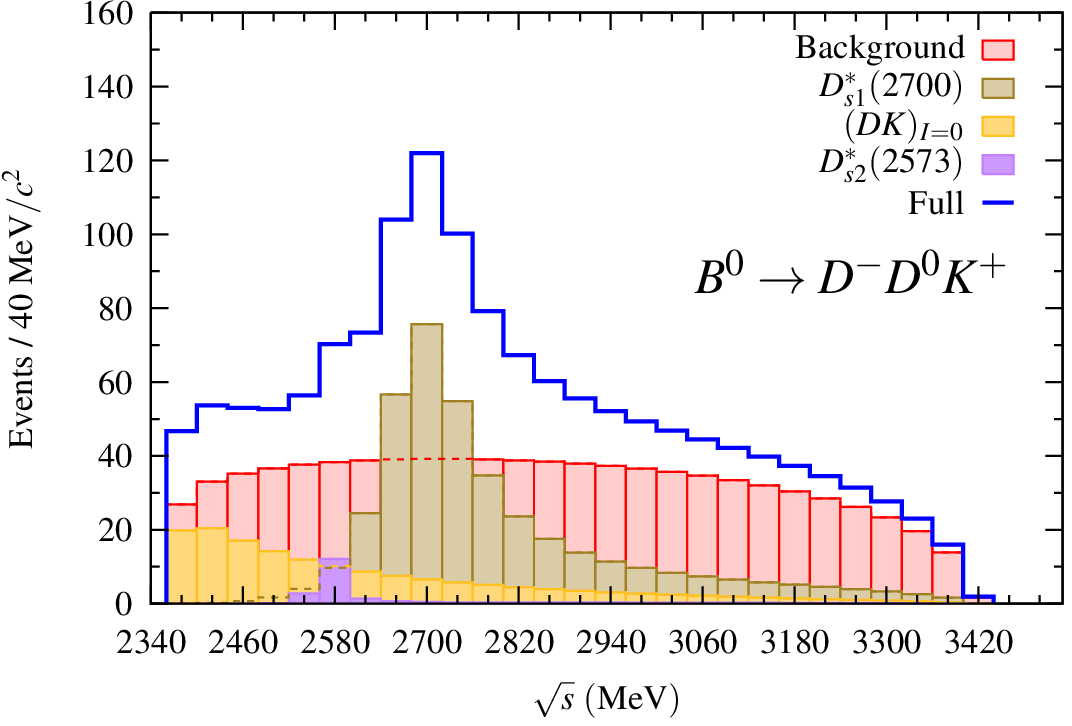} \\
\includegraphics[height=5.0cm,keepaspectratio]{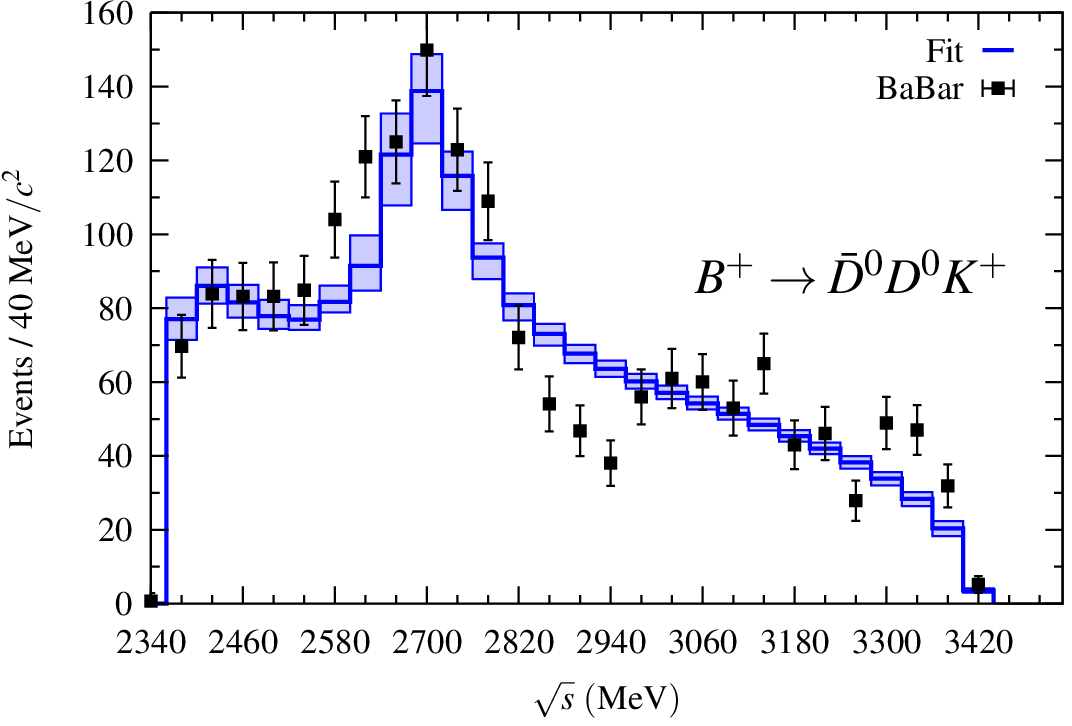} &
\includegraphics[height=5.0cm,keepaspectratio]{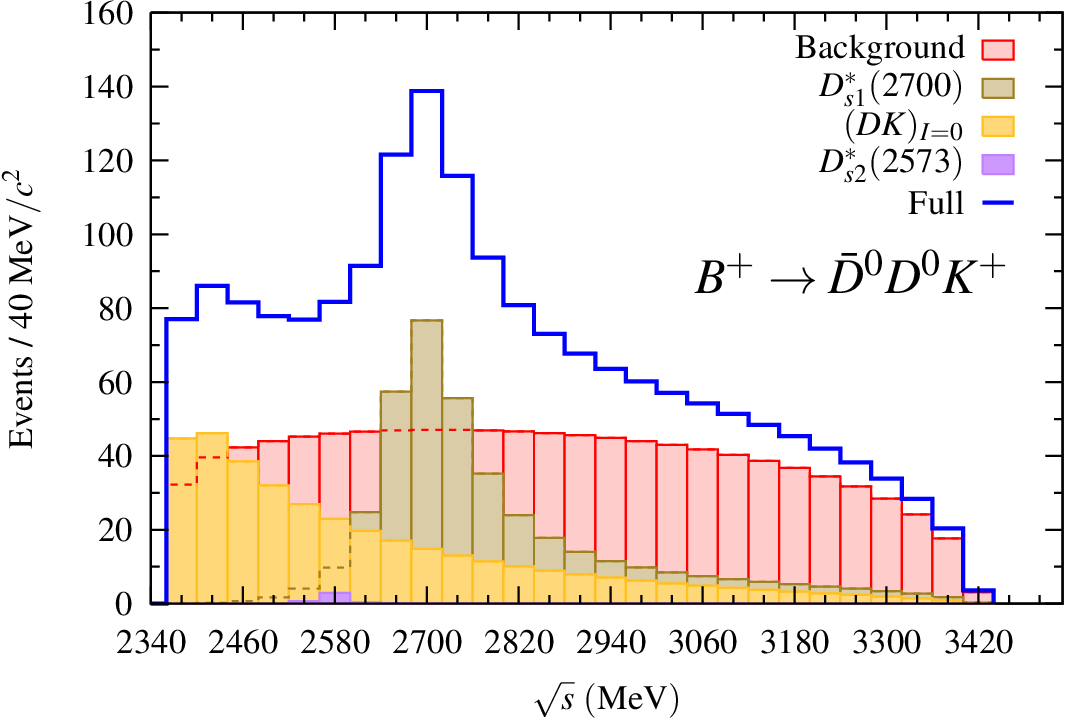}
\end{tabular}
\caption{The $D^0K^+$ invariant mass distributions for the reactions $B_s \to \pi^+ \bar{D}^0 K^-$ (top panels), $B^0 \to D^- D^0 K^+$ (middle panels) and $B^+ \to \xoverline{D}^0 D^0 K^+$ (lower panels). In the left panels, our fitted theoretical distributions (blue lines) together with its error (blue bands), and confronted with the experimental distributions (data taken from the LHCb \cite{Aaij:2014baa} and the BaBar \cite{Lees:2014abp} collaborations). In the right panels we show the different contributions to each decay. The fit shown here is the one called ``combined'' in Table~\ref{tab:results}.\label{fig:results}}
\end{figure*}

We start by performing two different fits to the LHCb and BaBar data separately. Among all the free parameters, we only show in Table~\ref{tab:results} the value of the one that is directly relevant for the $DK$ $T$-matrix, namely the subtraction constant $a(\mu)$. Alongside this value we also show the computed quantities stemming from each fit, $M_{D^\ast_{s0}}$, $P_{DK}$, $a_0$, and also the value $\chi^2/\text{d.o.f.}$. It can be seen that both fits have a good and similar quality, with $\chi^2/\text{d.o.f.} \simeq 1.3-1.4$, and that the values of the aforementioned quantities are compatible already at the $1\sigma$ level. The difference in the $D^\ast_{s0}(2317)$ mass in both fits is around $20\ \text{MeV}$, and the PDG \cite{Agashe:2014kda} average value, $2318 \pm 1\ \text{MeV}$, is comprised in the ranges obtained from both fits. Hence we perform a combined fit, also shown in Table~\ref{tab:results}, and the resulting mass, $M_{D^\ast_{s0}} = 2315^{+12}_{-17}\ ^{+10}_{-5}\ \text{MeV}$, is closer to the central value given by the PDG \cite{Agashe:2014kda} (albeit our errors are larger). For this combined fit, we show the mass spectra for the three reactions in Fig.~\ref{fig:results}. The enhancement at threshold is due to the $0(0^+)$ $DK$ amplitude, where the $D^\ast_{s0}(2317)$ appears as a pole, and the enhancement is then clearly explained by the presence of this bound state. The threshold enhancement is more clearly seen in the LHCb data (since it has a smaller bin size), where the $0(0^+)$ $DK$ amplitude dominates at threshold. In our analysis, the contribution of the latter amplitude is larger than that attributed to the $D^\ast_{s0}(2317)$ state by the LHCb analysis \cite{Aaij:2014baa} in the $B_s \to \pi^- \bar{D}^0 K^-$ amplitude. On the contrary, it can be seen that the contributions of this amplitude to the distributions in the two BaBar reactions, $B^0 \to D^- D^0 K^+$ and $B^+ \to \xoverline{D}^0 D^0 K^+$, is similar to that reported in the BaBar experimental analysis \cite{Lees:2014abp}, although it is attributed there to an {\it exponential background}, similarly as done by the Belle collaboration \cite{Brodzicka:2007aa}.

Surprisingly enough, from these experiments we can learn not only about the mass of the $D^\ast_{s0}(2317)$, but also about its nature, namely, about its $DK$ component $P_{DK}$, computed by means of Eq.~\eqref{eq:prob}. For the separate fits to the LHCb and BaBar data we get $P_{DK} = 74^{+7}_{-6}\ ^{+9}_{-1}\%$ and $67^{+5}_{-7}\ ^{+6}_{-10}\%$, respectively, whereas the combined fit gives $P_{DK} = 70^{+4}_{-6}\ ^{+4}_{-8}\%$. This result is similar to that obtained in Ref.~\cite{sasa3}. This large $DK$ component implies a mostly $DK$ molecular nature of $D^\ast_{s0}(2317)$.

\begin{table*}[t]
\begin{tabular}{cllccc} \hline
Fit      & $M_{D^\ast_{s0}}\ \text{(MeV)}$  & $P_{DK}(\%)$        & $a_0\ \text{(fm)}$      & $a(\mu)$                & $\chi^2/\text{d.o.f.}$ \\ \hline\hline
LHCb     & $2326^{+16}_{-16}\ ^{+1}_{-5}$  & $74^{+7}_{-6}\ ^{+9}_{-1}$  & $-1.10^{+0.19}_{-0.39}\ ^{+0.01}_{-0.02}$ & $-1.09^{+0.12}_{-0.10}\ ^{+0.01}_{-0.04}$ & $1.43$ \\
BaBar    & $2306^{+14}_{-23}\ ^{+16}_{-9}$ & $67^{+5}_{-7}\ ^{+6}_{-10}$ & $-0.87^{+0.15}_{-0.15}\ ^{+0.11}_{-0.18}$ & $-1.21^{+0.09}_{-0.13}\ ^{+0.10}_{-0.06}$ & $1.32$ \\
Combined & $2315^{+12}_{-17}\ ^{+10}_{-5}$ & $70^{+4}_{-6}\ ^{+4}_{-8}$  & $-0.95^{+0.15}_{-0.15}\ ^{+0.08}_{-0.13}$ & $-1.16^{+0.08}_{-0.10}\ ^{+0.06}_{-0.03}$ & $1.37$ \\ \hline
\end{tabular}
\caption{Mass and $DK$ probability of the $D^\ast_{s0}(2317)$ state, the $0(0^+)$ $DK$ scattering length and the fitted parameter $a(\mu)$, together with the reduced $\chi^2$, for each of the fits performed in this work. The first error is statistical, whereas the second one is systematic.\label{tab:results}}
\end{table*}

\section{Conclusions}\label{sec:conclusions}
We have performed a study of the $D^0 K^+$ invariant mass distributions for the weak decays $B_s \to \pi^+ \xoverline{D}^0 K^-$, $B^+ \to \xoverline{D}^0 D^0 K^+$ and $B^0 \to D^- D^0 K^+$, recently measured by the BaBar and LHCb collaborations \cite{Lees:2014abp,Aaij:2014baa}. In the three reactions, a clear enhancement at the $DK$ threshold is seen, which is difficult to interpret. The LHCb partly attributes this enhancement to the $D^\ast_{s0}(2317)$ resonance, but it is not a significant signal in their analysis. The BaBar collaboration models this enhancement through an exponential background, since they cannot draw definitive conclusions about its nature. In this work, we have shown that these enhancements are naturally explained by means of the $0(0^+)$ $DK$ elastic unitary amplitude, built from general principles (unitarity and HMChPT). This amplitude depends on a single parameter (a subtraction constant) fitted so as to reproduced the experimental distributions. A pole is found in this amplitude at a mass $M_{D^\ast_{s0}} = 2315^{+12}_{-17}\ ^{+10}_{-5}\ \text{MeV}$, which agrees with the PDG average value, although our errors are larger. Finally, by means of the Weinberg compositeness condition, we are also able to determine its $DK$ molecular nature, finding a $DK$ component $P_{DK} = 70^{+4}_{-6}\ ^{+4}_{-8}\ \%$.

\begin{acknowledgements}
M.~A. acknowledges financial support from the ``Juan de la Cierva'' program
(27-13-463B-731) from the Spanish MINECO. This work is supported in part by the
Spanish MINECO and European FEDER funds under the contracts FIS2014-51948-C2-1-P, FIS2014-51948-C2-2-P, FIS2014-57026-REDT and SEV-2014-0398, and by Generalitat
Valenciana under contract PROMETEOII/2014/0068. This work was partially supported by Open Partnership with Spain of JSPS Bilateral Joint Research Projects, and the work of DJ was supported by Grants-in-Aid for Scientific Research (No. 25400254).
\end{acknowledgements}

\bibliographystyle{plain}

\end{document}